\documentclass[12pt,a4paper]{article}
\usepackage{graphicx}
\title{Mathematical model of a flow of reacting substances in a channel of network}
\author{Nikolay K. Vitanov$^1$, Kaloyan N. Vitanov$^1$, Zlatinka I. Dimitrova$^2$}
\date{$^1$Institute of Mechanics, Bulgarian Academy of Sciences, Acad. G. Bonchev Str.,
	Bl. 4, 1113 Sofia, Bulgaria \\
$^2$	"G. Nadjakov" Institute of Solid State Physics, Bulgarian Academy of Sciences, Blvd. Tzarigradsko Chaussee 72, 1172 Sofia, Bulgaria }
\begin{document}
\maketitle
\begin{abstract}
Complex systems often have features that can be modeled by
advanced mathematical tools [1]. Of special interests are the features 
of complex systems that have a network structure as such systems are 
important for modeling technological and social processes [3, 4]. 
In our previous research we have discussed the flow of a single substance 
in a channel of network. It may happen however that two substances flow 
in the same channel of network. In addition the substances may react and
then the question arises about the distribution of the amounts of the 
substances in the segments of the channel. A study of the dynamics of 
the flow of the substances as well as a study of the distribution of 
the substances is presented in this paper on the base of a discrete - 
time model of flow of substances in the nodes of a channel of a network. 
\end{abstract}
\section{Introduction}
Complex systems are often modeled by means of tools from theories of nonlinear
dynamics, nonlinear time series analysis and nonlinear partial differential equations
\cite{d1} - \cite{dx}. The development of minimal cost transportation models  \cite{ff} stimulated 
much the research on network flows. The studies spread fast to the 
problems, e.g.,  for searching of:  minimal number of individuals to meet 
a fixed schedule; possible maximal flows in a network; optimum facility 
layout and location \cite{ch1}; optimal electronic route guidance in urban 
traffic networks \cite{hani},  etc. \cite{gomori} - \cite{boz}.
Below we shall consider the problem of motion of a substance
through a network channel in presence of possibility for "leakage" in the
nodes of the channel: the substance may be lost or used  in  some process. 
One application of the model is for description of the flow of some  
substance through a channel with use of part of substance for some 
industrial process in the nodes of the channel. The model has also another 
application: studying  large human  migration flows. 
Imagine a chain of countries that form a migration channel. The first
country of this channel may have a sea border and the migrants may come to this
country (called entry country of the channel) through this sea border.
In addition one or several countries of the channel may be preferred by
migrants. Such choice of an illustration of the model is motivated also by
actuality of the problem of human migration  \cite{everet} especially
after the large migration  flows directed to  Europe in  2015. 
Human migration models are of interest also for applied mathematics
as they can be classified as probability models, Markov chain models of 
migration \cite{will99}- \cite{brl08}) or deterministic models 
\cite{grd05}, \cite{z1}, \cite{el1}. Human migration is closely connected to 
migration networks \cite{fawcet}, to ideological struggles 
\cite{vit1}, \cite{vit2}  and to waves and statistical distributions in 
population systems \cite{vit3} -\cite{vit6}. 
\par 
The paper is organized as follows. In Sect.2 we discuss a model for motion of
two reacting substances in a channel containing finite number of nodes.
In Sect. 3 we derive statistical distributions  for the 
amounts of substances in the nodes of the channel are obtained. 
Several concluding remark remarks are summarized in Sect. 4.

\section{Mathematical formulation of the model}
Let us consider a large enough network consisting of nodes connected by 
edges. We assume the existence of a channel in this network. The channel 
consists  of a chain of $N+1$ nodes (labeled  from $0$ to $N$) connected 
by corresponding edges. Each edge connects two nodes and each node is 
connected to two edges except for the $0$-th node and $N$-th node that 
are connected by one edge. We assume that two substances $X$ and $Y$ can 
move through the channel. The substances enter the channel through
the $0$-th node and move to the nodes labeled by increasing number. The 
time is discrete and consists of equal time
intervals. At each time interval the substance in a node of the channel 
can participate in one of the following four processes: (a) the substance 
remains in the same cell and stays in the channel (i.e., there is no leak 
of the substance from the channel); (b) the substance moves to the next 
cell (i.e., the substance moves from the node $m$ to the node $m+1$); 
(c) the substance "leaks" from the channel: this means that the substance 
do not belong anymore to the channel. Such substance may stay in the 
corresponding node of the channel or may spread through the
network; (d) substances $X$ and $Y$ react and as a result some amount of 
substance $Z$ is created. The substance $Z$ can move along the nodes of
the channel too (in the same direction as the substances $X$ and $Y$). 
\par
Let us formalize mathematically the above considerations. We denote the 
amounts of substances $X,Y,Z$ as $x,y,z$ respectively. The following 
processes can be observed in a node of the channel: (i) exchange (inflow) 
of basic sybstances $x$ and $y$ from the environment to the $0$-th node 
of the channel; (ii) exchange (inflow) of substances $x,y,z$ with the 
previous node of the channel (for the nodes $1,\dots,N$-th  of the channel);
(iii) exchange (outflow) of substances $x,y,z$ with the next node of the 
channel (for the nodes $0,\dots,N-1$ of the channel); (iv) "leakages":  
exchange (outflow) of substances $x,y,z$ with the correspondent node of 
the network; (v) interaction between the two substances $X$ and $Y$ that 
leads to occurring of the third substance $Z$.
\par
We consider discrete time $t_k$, $k=0,1,2,\dots$ as in the case 
\cite{vr2}. Let us denote the amount of
the substances in the $i$-th node of the channel at the beginning of the
time interval $[t_k, t_k + \Delta t]$ as $x_i(t_k)$, $y_i(t_k)$ and $z_i(t_k)$. 
For the processes happening in this time interval in the $n$-th node of 
the channel we shall use the following  notations: 
$xi^e_n(t_k)$, $yi^e_n(t_k)$, $zi^e_n(t_k)$  and $xo^e_n(t_k)$, $yo^e_n(t_k)$, 
$zo^e_n(t_k)$ are the amounts of inflow and outflow  of substance from the 
environment to the $n$-th node  of the channel (the upper index $e$ denotes 
that the quantities are for the environment);  $xo^c_n(t_k)$, 	$yo^c_n(t_k)$, 	
$zo^c_n(t_k)$ are the amounts of outflow of substances from the $n$-th node of the
channel to the $(n+1)$-th node of the channel (the upper index $c$ denotes 
that the quantities are for the channel); $xi^c_n(t_k)$, $yi^c_n(t_k)$, 
$zi^c_n(t_k)$ are the amounts of the inflow of substance from the $(n+1)$ node of 
the channel to the $n$-th node of the channel; $xo^n_n(t_k)$, $yo^n_n(t_k)$, 
$zo^n_n(t_k)$ and $xi^n_n(t_k)$, $yi^n_n(t_k)$, $zi^n_n(t_k)$ are the amounts 
of outflow  and inflow of substance between the $n$-th node of the channel and  
the corresponding node of the coupled network (the upper index $n$ denotes 
that the quantities are for the network). $xv_n(t_k)$, $yv_n(t_k)$ are the 
amounts of substances $x$ and $y$ that interact in the $n$-th node of the 
channel and as a result the substance $zc_n(t_k)$ is created in the $n$-th 
node of the channel. 
\par
We shall assume that: (i) there is no inflow of 
substance from the nodes of the network to the channel; (ii) there is no 
outflow of substance from the $0$-th node of the channel to the environment;
(iii) there is no inflow of substance from the $i$-th node of the channel to the
$i-1$-th node of the channel, $i=1,\dots,N$.
For the particular case described above the system of model equations becomes
($i=1,\dots,N-1$)
	\begin{eqnarray}\label{e1}
	x_0(t_{k+1}) &=& x_0(t_k)+ xi^e_0(t_k)  - xo^c_0(t_k)  -xo^n_0(t_k) -xv_0(t_k) \nonumber \\
	y_0(t_{k+1}) &=& y_0(t_k)+ yi^e_0(t_k)  - yo^c_0(t_k)  -yo^n_0(t_k) -yv_0(t_k)\nonumber \\
	z_0(t_{k+1}) &=& 	z_0(t_k) + xv_0(t_k) + yv_0(t_k) - zo^c_0(t_k)  -zo^n_0(t_k) \nonumber \\
	x_i(t_{k+1}) &=& x_i(t_k)+xo^c_{i-1}(t_k) - xo^c_i(t_k)  - xo^n_i(t_k) -xv_i(t_k), \nonumber \\
	y_i(t_{k+1}) &=& y_i(t_k)+yo^c_{i-1}(t_k) - yo^c_i(t_k)  - yo^n_i(t_k) -yv_i(t_k),\nonumber \\
	z_i(t_{k+1}) &=& xv_i(t_k) + yv_i(t_k) + zo^c_{i-1}(t_k) - zo^c_i(t_k)  - zo^n_i(t_k)\nonumber \\
	x_N(t_{k+1}) &=& x_N(t_k) + xo^c_{N-1}(t_k) - xo^n_N(t_k) - xv_N(t_k) 
	\nonumber \\
	y_N(t_{k+1}) &=& y_N(t_k) + yo^c_{N-1}(t_k) - yo^n_N(t_k) - yv_N(t_k) \nonumber \\
	z_N(t_{k+1}) &=& xv_N(t_k) + yv_N(t_k) + zo^c_{N-1}(t_k) - zo^n_N(t_k)
	\end{eqnarray}
	\par
	Below we shall study the following particular cases of the quantities from
	the system of equations (\ref{e1}) ($i=2,\dots, N-2$)
	\begin{eqnarray}\label{pc}
	xi^e_0(t_k) &=& \sigma_x(t_k) x_0(t_k); \ \ xo^c_0(t_k) = f_{x,0}(t_k) x_0(t_k);
	\nonumber \\
	yi^e_0(t_k) &=& \sigma_y(t_k) y_0(t_k); \ \ yo^c_0(t_k) = f_{y,0}(t_k) y_0(t_k);
	\nonumber \\
	xo^n_0(t_k) &=& \gamma_{x,0}(t_k) x_0(t_k); \ \ xv_0(t_k) = \omega_0(t_k) x_0(t_k) y_0(t_k)	
	\nonumber \\
	yo^n_0(t_k) &=& \gamma_{y,0}(t_k) y_0(t_k); \ \ yv_0(t_k) = \pi_0(t_k) x_0(t_k) y_0(t_k)	
	\nonumber \\
	zo^c_0(t_k) &=& f_{z,0}(t_k) z_0(t_k); \ \ zo^n_0(t_k) = \gamma_{z,0}(t_k) z_0(t_k)
	\nonumber \\
	 xo^c_{i-1}(t_k) &=& f_{x,i-1}(t_k) x_{i-1}(t_k); \ \
	 xo^n_{i-1}(t_k) = \gamma_{x,i-1}(t_k) x_{i-1}(t_k);
	 \nonumber \\  
	 yo^c_{i-1}(t_k) &=& f_{y,i-1}(t_k) y_{i-1}(t_k); 
	 yo^n_{i-1}(t_k) = \gamma_{y,i-1}(t_k) y_{i-1}(t_k);
	 \nonumber \\ 
	 zo^c_{i-1}(t_k) &=& f_{z,i-1}(t_k) z_{i-1}(t_k); \ \
	 zo^n_{i-1}(t_k) = \gamma_{z,i-1}(t_k) z_{i-1}(t_k);
	\nonumber \\
	xo^c_{i}(t_k) &=& f_{x,i}(t_k) x_i(t_k); \ \ xo^n_{i}(t_k) = \gamma_{x,i}(t_k) x_i(t_k); \nonumber \\
	yo^c_{i}(t_k) &=& f_{y,i}(t_k) y_i(t_k); \ \ yo^n_{i}(t_k) = \gamma_{y,i}(t_k) y_i(t_k); \nonumber \\
	zo^c_{i}(t_k) &=& f_{z,i}(t_k) z_i(t_k); 
	 zo^n_{i}(t_k) = \gamma_{z,i}(t_k) z_i(t_k); \nonumber \\
	xv_{i-1}(t_k) &=& \omega_{i-1}(t_k) x_{i-1}(t_k) y_{i-1}(t_k) \ \ ; yv_{i-1}(t_k) = \pi_{i-1}(t_k) x_{i-1}(t_k) y_{i-1}(t_k)
	 \nonumber \\
	xv_{i}(t_k) &=& \omega_{i}(t_k) x_{i}(t_k) y_{i}(t_k) \ \ ; yv_{i}(t_k) = \pi_{i}(t_k) x_{i}(t_k) y_{i}(t_k)
	 \nonumber \\
	xo^c_{N-1}(t_k) &=& f_{x,N-1}(t_k) x_{N-1}(t_k); \ \ 
	xo^n_{N-1}(t_k) = \gamma_{x,N-1}(t_k) x_{N-1}(t_k); \nonumber \\
	yo^c_{N-1}(t_k) &=& f_{y,N-1}(t_k) y_{N-1}(t_k); \ \ 
	yo^n_{N-1}(t_k) = \gamma_{y,N-1}(t_k) y_{N-1}(t_k);
	\nonumber \\
	zo^c_{N-1}(t_k) &=& f_{z,N-1}(t_k) z_{N-1}(t_k); \ \ 
	zo^n_{N-1}(t_k) = \gamma_{z,N-1}(t_k) z_{N-1}(t_k);
	\nonumber \\
	xo^n_{N}(t_k) &=& \gamma_{x,N}(t_k) x_N(t_k); \ \
	xv_N(t_k) = \omega_N(t_k) x_N(t_k) y_N(t_k)	\nonumber \\
	yo^n_{N}(t_k) &=& \gamma_{y,N}(t_k) y_N(t_k); \ \
	yv_N(t_k) = \pi_N(t_k) x_N(t_k) y_N(t_k)	\nonumber \\
	zo^n_{N}(t_k) &=& \gamma_{z,N}(t_k) z_N(t_k); \ \
	\end{eqnarray}
	For this particular case the system of equations (\ref{e1}) becomes
	($i=1,\dots,N-1$)
	\begin{eqnarray}\label{ex1}
	x_0(t_{k+1}) &=& x_0(t_k)+ \sigma_x(t_k) x_0(t_k) -
	f_{x,0}(t_k) x_0(t_k) - \gamma_{x,0}(t_k) x_0(t_k) -
	\omega_0(t_k) x_0(t_k) y_0(t_k), 
\nonumber \\	
	y_0(t_{k+1}) &=& y_0(t_k)+ \sigma_y(t_k) y_0(t_k) -
	f_{y,0}(t_k) y_0(t_k) - \gamma_{y,0}(t_k) y_0(t_k) -
	\pi_0 (t_k) x_0(t_k) y_0(t_k),
\nonumber \\
	z_0(t_{k+1}) &=& 	z_0(t_k) + \omega_0(t_k) x_0(t_k) y_0(t_k) + \pi_0(t_k) x_0(t_k) y_0(t_k) 
	- f_{z,0}(t_k) z_0(t_k) - \gamma_{z,0}(t_k) z_0(t_k).
\nonumber \\
x_i(t_{k+1}) &=& x_i(t_k) + f_{x,i-1}(t_k) x_{i-1}(t_k) -
f_{x,i}(t_k) x_i(t_k) \gamma_{x,i}(t_k) x_i(t_k) -
\omega_{i}(t_k) x_{i}(t_k) y_{i}(t_k), \nonumber \\
y_i(t_{k+1}) &=& y_i(t_k) + f_{y,i-1}(t_k) y_{i-1}(t_k) - 
f_{y,i}(t_k) y_i(t_k) - \gamma_{y,i}(t_k) y_i(t_k) -
\pi_{i}(t_k) x_{i}(t_k) y_{i}(t_k),
\nonumber \\
z_i(t_{k+1}) &=&  z_i(t_k) + 
\omega_{i}(t_k) x_{i}(t_k) y_{i}(t_k) + \pi_{i}(t_k) x_{i}(t_k) y_{i}(t_k) +f_{z,i-1}(t_k) z_{i-1}(t_k) - 
\nonumber \\
&& f_{z,i}(t_k) z_i(t_k) - \gamma_{z,i}(t_k) z_i(t_k).
\nonumber \\
x_N(t_{k+1}) &=& x_N(t_k) +
f_{x,N-1}(t_k) x_{N-1}(t_k) - \gamma_{x,N}(t_k) x_N(t_k) -
\omega_N(t_k) x_N(t_k) y_N(t_k),
\nonumber \\
y_N(t_{k+1}) &=& y_N(t_k) +
f_{y,N-1}(t_k) y_{N-1}(t_k) - \gamma_{y,N}(t_k) y_N(t_k) - 
\pi_N(t_k) x_N(t_k) y_N(t_k),
 \nonumber \\
z_N(t_{k+1}) &=&
 \omega_N(t_k) x_N(t_k) y_N(t_k) + \pi_N(t_k) x_N(t_k) y_N(t_k) +  f_{z,N-1}(t_k) z_{N-1}(t_k) - \nonumber \\
 && \gamma_{z,N}(t_k) z_N(t_k) \nonumber \\
	\end{eqnarray}
We shall study the model equations (\ref{ex1}) in more detail below.
\section{Case of constant values of the parameters and stationary state 
of functioning of the channel}
Analytical results can be obtained for the model described by Eqs.(\ref{ex1}) 
for the case when the parameters of the model are time independent.
Then the system of model equations becomes ($i=1,\dots,N-1$)
\begin{eqnarray}\label{ey1}
x_0(t_{k+1}) &=& x_0(t_k)+ \sigma_x x_0(t_k) -
f_{x,0} x_0(t_k) - \gamma_{x,0} x_0(t_k) -
\omega_0 x_0(t_k) y_0(t_k), 
\nonumber \\	
y_0(t_{k+1}) &=& y_0(t_k)+ \sigma_y y_0(t_k) -
f_{y,0} y_0(t_k) - \gamma_{y,0} y_0(t_k) -
\pi_0  x_0(t_k) y_0(t_k),
\nonumber \\
z_0(t_{k+1}) &=& 	z_0(t_k) + \omega_0 x_0(t_k) y_0(t_k) + \pi_0 x_0(t_k) y_0(t_k) 
- f_{z,0} z_0(t_k) - \gamma_{z,0} z_0(t_k).
\nonumber \\
x_i(t_{k+1}) &=& x_i(t_k) + f_{x,i-1} x_{i-1}(t_k) -
f_{x,i} x_i(t_k) - \gamma_{x,i} x_i(t_k) -
\omega_{i} x_{i}(t_k) y_{i}(t_k), \nonumber \\
y_i(t_{k+1}) &=& y_i(t_k) + f_{y,i-1} y_{i-1}(t_k) - 
f_{y,i} y_i(t_k) - \gamma_{y,i} y_i(t_k) -
\pi_{i} x_{i}(t_k) y_{i}(t_k),
\nonumber \\
z_i(t_{k+1}) &=&  z_i(t_k) + 
\omega_{i} x_{i}(t_k) y_{i}(t_k) + \pi_{i} x_{i}(t_k) y_{i}(t_k) +f_{z,i-1} z_{i-1}(t_k) - 
\nonumber \\
&& f_{z,i} z_i(t_k) - \gamma_{z,i} z_i(t_k).
\nonumber \\
x_N(t_{k+1}) &=& x_N(t_k) +
f_{x,N-1} x_{N-1}(t_k) - \gamma_{x,N} x_N(t_k) -
\omega_N x_N(t_k) y_N(t_k),
\nonumber \\
y_N(t_{k+1}) &=& y_N(t_k) +
f_{y,N-1} y_{N-1}(t_k) - \gamma_{y,N} y_N(t_k) - 
\pi_N x_N(t_k) y_N(t_k),
\nonumber \\
z_N(t_{k+1}) &=& z_N(t_k) + 
\omega_N x_N(t_k) y_N(t_k) + \pi_N x_N(t_k) y_N(t_k) +  f_{z,N-1} z_{N-1}(t_k) - \nonumber \\
&& \gamma_{z,N} z_N(t_k) \nonumber \\
\end{eqnarray}
	\par
Let us remember the distribution of substance in the cells of the channel for tha case of stationary state of flow in the channel in presence of one substance $x$. In this case $x_i(t_k) = x_i^*$. which arises when $x_i(t_{k+1})=x_i(t_k)$ (i.e., there is a motion of substance through the cells of the channel but the motion happens in such a way that the amount of the substance in a given cell remains the same for the following time intervals). For this case
$\sigma = f_{0} + \gamma_{0}$ (the substance that enters the channel moves to the next cells or leaks) and $x^*$ is a free parameter. This is not the case of two interacting substances.  For the stationary case we obtain from  Eqs.(\ref{ey1}) 
\begin{eqnarray}\label{ez1}
x_0^* &=& \frac{\sigma_y - f_{y,0} - \gamma_{y,0}}{\pi_0}, \ \ y_0^* =  \frac{\sigma_x -f_{x,0} - \gamma_{x,0} }{\omega_0}
\nonumber\\ 
z_0^* &=& \left(\frac{1}{\omega_0} + \frac{1}{\pi_0}\right) \frac{(\sigma_x - f_{x,0} - \gamma_{x,0})(\sigma_y - f_{y,0} - \gamma_{y,0})}{ f_{z,0} + \gamma_{z,0}} ,
\nonumber \\
\end{eqnarray}	
i.e., stationary regime of functioning of the channel is possible only if the amounts of substances in the entry cell
of the channel have the values selected by Eqs.(\ref{ez1}).
We can proceed and obtain analytical results for the distributions of the substances $x_i^*$, $y_i^*$ and $z_i^*$
in the cells of the channel. If the number of cells is
large however the relationships for these  distributions become very large after just several values of increasing parameter $i$. Thus in order to illustrate our analytical results we shall consider the particular case where the
substance $y$ is a catalysts of converting of part of substance $x$ into the substance $z$. In this case
$\pi_i = 0$, $i=0,\dots, N$ and the model system of equations
(stationary regime, constant values of the parameters)
becomes ($i=1,\dots,N-1$)
\begin{eqnarray}\label{er1}
0&=&  \sigma_x x_0^* -
f_{x,0} x_0^* - \gamma_{x,0} x_0^* - \omega_0 x_0^* y_0^*, 
\nonumber \\	
0 &=&  \sigma_y y_0^* - f_{y,0} y_0^* - \gamma_{y,0} y_0^*,
\nonumber \\
0 &=&  \omega_0 x_0^* y_0^*  - f_{z,0} z_0^* - \gamma_{z,0} z_0^*.
\nonumber \\
0&=& f_{x,i-1} x_{i-1}^* -f_{x,i} x_i^* - \gamma_{x,i} x_i^* -
\omega_{i} x_{i}^* y_{i}^*, \nonumber \\
0 &=&  f_{y,i-1} y_{i-1}^* - f_{y,i} y_i^* - \gamma_{y,i} y_i^*,
\nonumber \\
0 &=&   
\omega_{i} x_{i}^* y_{i}^* +f_{z,i-1} z_{i-1}^* - 
f_{z,i} z_i^* - \gamma_{z,i} z_i^*.
\nonumber \\
0 &=& f_{x,N-1} x_{N-1}^* - \gamma_{x,N} x_N^* -
\omega_N x_N^* y_N^*,
\nonumber \\
0 &=& f_{y,N-1} y_{N-1}^* - \gamma_{y,N} y_N^*,
\nonumber \\
0&=&
\omega_N x_N^* y_N^* +  f_{z,N-1} z_{N-1}^* - \gamma_{z,N} z_N^* 
\end{eqnarray}
From Eqs. (\ref{er1})  we obtain that $y_0^*$
is free parameter (for the entry cell $\sigma_y = f_{y,0} + \gamma_{y,0}$, 
i.e., the substance that enters the cell leaves it or leaks and the 
total amount of substance in the cell do not change). In addition	
\begin{eqnarray}\label{yy1}
y_i^* = y_0^* \prod \limits_{j=1}^i \frac{f_{y,j-1}}{f_{y,j} + \gamma_{y,j}}; \ \
y_N^* = y_0^* \frac{f_{y,N-1}}{\gamma_{y,N}}\prod \limits_{j=1}^{N-1} \frac{f_{y,j-1}}{f_{y,j}+\gamma_{y,j}}
\end{eqnarray}
The total amount of the substance $y$ in the channel is
\begin{equation}\label{yy2}
y^* = y_0^* \Bigg[ 1 + \sum \limits_{k=1}^{N-1} \prod \limits_{j=1}^k \frac{f_{y,j-1}}{f_{y,j} + \gamma_{y,j}}
+ \frac{f_{y,N-1}}{\gamma_{y,N}}\prod \limits_{j=1}^{N-1} \frac{f_{y,j-1}}{f_{y,j}+\gamma_{y,j}}
\Bigg]
\end{equation}	
We  consider the statistical distribution of the amount of substance along the nodes of the channel $\xi^*_i=y_i^*/y^*$. For this distribution we obtain
\begin{eqnarray}\label{dstr_main1}
\xi^*_0 &=& \frac{1}{\Bigg[ 1 + \sum \limits_{k=1}^{N-1} \prod \limits_{j=1}^k \frac{f_{y,j-1}}{f_{y,j} + \gamma_{y,j}}
	+ \frac{f_{y,N-1}}{\gamma_{y,N}}\prod \limits_{j=1}^{N-1} \frac{f_{y,j-1}}{f_{y,j}+\gamma_{y,j}}
	\Bigg]} \nonumber \\
\xi_i^* &=& \frac{\prod \limits_{j=1}^i \frac{f_{y,j-1}}{f_{y,j}+\gamma_{y,j}}}{\Bigg[ 1 + \sum \limits_{k=1}^{N-1} \prod \limits_{j=1}^k \frac{f_{y,j-1}}{f_{y,j} + \gamma_{y,j}}
	+ \frac{f_{y,N-1}}{\gamma_{y,N}}\prod \limits_{j=1}^{N-1} \frac{f_{y,j-1}}{f_{y,j}+\gamma_{y,j}}
	\Bigg]}; i=1,\dots,N-1 \nonumber \\
\xi_N^* &=& \frac{\frac{f_{y,N-1}}{\gamma_{y,N}}\prod \limits_{j=1}^{N-1}
	\frac{f_{y,j-1}}{f_{y,j}+\gamma_{y,j}}}{\Bigg[ 1 + \sum \limits_{k=1}^{N-1} \prod \limits_{j=1}^k \frac{f_{y,j-1}}{f_{y,j} + \gamma_{y,j}}
	+ \frac{f_{y,N-1}}{\gamma_{y,N}}\prod \limits_{j=1}^{N-1} \frac{f_{y,j-1}}{f_{y,j}+\gamma_{y,j}}
	\Bigg]}
\end{eqnarray}	
$x_0^*$ is a free parameter in the case 
of the presence of the following relationship
among the parameters for the substance $x$ in the entry cell
and the parameter  $y_0^*$
\begin{equation}\label{xx1}
\sigma_x = f_{x,0} + \gamma_{x,0} + \omega_0 y_0^*
\end{equation}
Then
\begin{eqnarray}\label{xx2}
x_i^* = x_0^* \prod \limits_{j=1}^i \frac{f_{x,j-1}}{f_{x,j} + \gamma_{x,j} +  \omega_i y_0^* \prod \limits_{j=1}^i \frac{f_{y,j-1}}{f_{y,j} + \gamma_{y,j}}}; \nonumber \\
x_N^* = x_0^* \frac{f_{x,N-1}}{\gamma_{x,N} + \omega_N y_0^* \frac{f_{y,N-1}}{\gamma_{y,N}}\prod \limits_{j=1}^{N-1} \frac{f_{y,j-1}}{f_{y,j}+\gamma_{y,j}}}\prod \limits_{j=1}^{N-1} \frac{f_{x,j-1}}{f_{x,j}+\gamma_{x,j} +  \omega_i y_0^* \prod \limits_{j=1}^i \frac{f_{y,j-1}}{f_{y,j} + \gamma_{y,j}}}	
\end{eqnarray}
The total amount of the substance $x$ in the channel is
\begin{eqnarray}\label{xx3}
x^* = \Bigg[ 1+ \sum \limits_{k=1}^{N-1}  \prod \limits_{j=1}^k \frac{f_{x,j-1}}{f_{x,j} + \gamma_{x,j} +  \omega_i y_0^* \prod \limits_{j=1}^k \frac{f_{y,j-1}}{f_{y,j} + \gamma_{y,j}}}  + \nonumber \\
 \frac{f_{x,N-1}}{\gamma_{x,N} + \omega_N y_0^* \frac{f_{y,N-1}}{\gamma_{y,N}}\prod \limits_{j=1}^{N-1} \frac{f_{y,j-1}}{f_{y,j}+\gamma_{y,j}}}\prod \limits_{j=1}^{N-1} \frac{f_{x,j-1}}{f_{x,j}+\gamma_{x,j} +  \omega_i y_0^* \prod \limits_{j=1}^i \frac{f_{y,j-1}}{f_{y,j} + \gamma_{y,j}}} \Bigg]
\end{eqnarray}
The distribution $\zeta_i = x_i^*/x^*$ of the substance $x$
in the cells of the channel is
\begin{eqnarray} \label{xx4}
\zeta_0 &=& 1 \Bigg / \Bigg[ 1+ \sum \limits_{k=1}^{N-1}  \prod \limits_{j=1}^k \frac{f_{x,j-1}}{f_{x,j} + \gamma_{x,j} +  \omega_i y_0^* \prod \limits_{j=1}^k \frac{f_{y,j-1}}{f_{y,j} + \gamma_{y,j}}}  + \nonumber \\
&&\frac{f_{x,N-1}}{\gamma_{x,N} + \omega_N y_0^* \frac{f_{y,N-1}}{\gamma_{y,N}}\prod \limits_{j=1}^{N-1} \frac{f_{y,j-1}}{f_{y,j}+\gamma_{y,j}}}\prod \limits_{j=1}^{N-1} \frac{f_{x,j-1}}{f_{x,j}+\gamma_{x,j} +  \omega_i y_0^* \prod \limits_{j=1}^i \frac{f_{y,j-1}}{f_{y,j} + \gamma_{y,j}}} \Bigg] \nonumber \\
\zeta_i &=& x_0^* \prod \limits_{j=1}^i \frac{f_{x,j-1}}{f_{x,j} + \gamma_{x,j} +  \omega_i y_0^* \prod \limits_{j=1}^i \frac{f_{y,j-1}}{f_{y,j} + \gamma_{y,j}}}
\bigg / \Bigg[ 1+ \sum \limits_{k=1}^{N-1}  \prod \limits_{j=1}^k \frac{f_{x,j-1}}{f_{x,j} + \gamma_{x,j} +  \omega_i y_0^* \prod \limits_{j=1}^k \frac{f_{y,j-1}}{f_{y,j} + \gamma_{y,j}}}  + \nonumber \\
&&\frac{f_{x,N-1}}{\gamma_{x,N} + \omega_N y_0^* \frac{f_{y,N-1}}{\gamma_{y,N}}\prod \limits_{j=1}^{N-1} \frac{f_{y,j-1}}{f_{y,j}+\gamma_{y,j}}}\prod \limits_{j=1}^{N-1} \frac{f_{x,j-1}}{f_{x,j}+\gamma_{x,j} +  \omega_i y_0^* \prod \limits_{j=1}^i \frac{f_{y,j-1}}{f_{y,j} + \gamma_{y,j}}} \Bigg] \nonumber \\
\zeta_N &=& x_0^* \frac{f_{x,N-1}}{\gamma_{x,N} + \omega_N y_0^* \frac{f_{y,N-1}}{\gamma_{y,N}}\prod \limits_{j=1}^{N-1} \frac{f_{y,j-1}}{f_{y,j}+\gamma_{y,j}}}\prod \limits_{j=1}^{N-1} \frac{f_{x,j-1}}{f_{x,j}+\gamma_{x,j} +  \omega_i y_0^* \prod \limits_{j=1}^i \frac{f_{y,j-1}}{f_{y,j} + \gamma_{y,j}}}	
\Bigg / \Bigg[ 1+ \nonumber \\
&& \sum \limits_{k=1}^{N-1}  \prod \limits_{j=1}^k \frac{f_{x,j-1}}{f_{x,j} + \gamma_{x,j} +  \omega_i y_0^* \prod \limits_{j=1}^k \frac{f_{y,j-1}}{f_{y,j} + \gamma_{y,j}}}  + \nonumber \\
&&
\frac{f_{x,N-1}}{\gamma_{x,N} + \omega_N y_0^* \frac{f_{y,N-1}}{\gamma_{y,N}}\prod \limits_{j=1}^{N-1} \frac{f_{y,j-1}}{f_{y,j}+\gamma_{y,j}}}\prod \limits_{j=1}^{N-1} \frac{f_{x,j-1}}{f_{x,j}+\gamma_{x,j} +  \omega_i y_0^* \prod \limits_{j=1}^i \frac{f_{y,j-1}}{f_{y,j} + \gamma_{y,j}}} \Bigg]
\end{eqnarray}
For the substance $z$ we obtain
\begin{eqnarray}\label{zz1}
z_0^* &=& \frac{\omega_0 x_0^* y_0^*}{f_{z,0}+\gamma_{z,0}}
\nonumber \\
z_i^* &=& \frac{\omega_i x_0^* y_0^* \Bigg( \prod \limits_{j=1}^i \frac{f_{x,j-1}}{f_{x,j} + \gamma_{x,j} +  \omega_i y_0^* \prod \limits_{j=1}^i \frac{f_{y,j-1}}{f_{y,j} + \gamma_{y,j}}} \Bigg) \bigg( \prod \limits_{j=1}^i \frac{f_{y,j-1}}{f_{y,j} + \gamma_{y,j}} \bigg)}{f_{z,i} + \gamma_{z,i}} + \nonumber \\
&& f_{z,i-1} \Bigg \{\frac{\omega_{i-1} x_0^* y_0^* \Bigg( \prod \limits_{j=1}^{i-1} \frac{f_{x,j-1}}{f_{x,j} + \gamma_{x,j} +  \omega_i y_0^* \prod \limits_{j=1}^{i-1} \frac{f_{y,j-1}}{f_{y,j} + \gamma_{y,j}}} \Bigg) \bigg( \prod \limits_{j=1}^{i-1} \frac{f_{y,j-1}}{f_{y,j} + \gamma_{y,j}} \bigg)}{f_{z,i-1} + \gamma_{z,i-1}} + \nonumber \\
&& f_{z,i-2} \Bigg[\dots + f_{z,1} \frac{\omega_0 x_0^* y_0^*}{f_{z,0}+ \gamma_{z,0}}  \Bigg]
 \Bigg \} , i=1,2,\dots, N-1 \nonumber \\
z_N^* &=& \frac{\omega_N x_0^* y_0^*}{\gamma_{z,N}} \Bigg(x_0^* \frac{f_{x,N-1}}{\gamma_{x,N} + \omega_N y_0^* \frac{f_{y,N-1}}{\gamma_{y,N}}\prod \limits_{j=1}^{N-1} \frac{f_{y,j-1}}{f_{y,j}+\gamma_{y,j}}}\prod \limits_{j=1}^{N-1} \frac{f_{x,j-1}}{f_{x,j}+\gamma_{x,j} +  \omega_i y_0^* \prod \limits_{j=1}^i \frac{f_{y,j-1}}{f_{y,j} + \gamma_{y,j}}} \Bigg) \times \nonumber \\
&& \bigg(  y_0^* \frac{f_{y,N-1}}{\gamma_{y,N}}\prod \limits_{j=1}^{N-1} \frac{f_{y,j-1}}{f_{y,j}+\gamma_{y,j}}
\bigg)+ 
\frac{\omega_i x_0^* y_0^* \Bigg( \prod \limits_{j=1}^{N-2} \frac{f_{x,j-1}}{f_{x,j} + \gamma_{x,j} +  \omega_i y_0^* \prod \limits_{j=1}^{N-2} \frac{f_{y,j-1}}{f_{y,j} + \gamma_{y,j}}} \Bigg) \bigg( \prod \limits_{j=1}^{N-2} \frac{f_{y,j-1}}{f_{y,j} + \gamma_{y,j}} \bigg)}{f_{z,i} + \gamma_{z,i}} + \nonumber \\
&& f_{z,N-2} \Bigg \{\frac{\omega_{i-1} x_0^* y_0^* \Bigg( \prod \limits_{j=1}^{N-3} \frac{f_{x,j-1}}{f_{x,j} + \gamma_{x,j} +  \omega_i y_0^* \prod \limits_{j=1}^{N-3} \frac{f_{y,j-1}}{f_{y,j} + \gamma_{y,j}}} \Bigg) \bigg( \prod \limits_{j=1}^{N-3} \frac{f_{y,j-1}}{f_{y,j} + \gamma_{y,j}} \bigg)}{f_{z,i-1} + \gamma_{z,i-1}} + \nonumber \\
&& f_{z,N-3} \Bigg[\dots + f_{z,1} \frac{\omega_0 x_0^* y_0^*}{f_{z,0}+ \gamma_{z,0}}  \Bigg]
\Bigg \}
\end{eqnarray}
Thus for the distribution $\mu_i = z_i^*/z^*$ we have
\begin{equation}\label{zz2}
\mu_i = \frac{z_i^*}{z_0^* + \sum \limits_{k=1}^{N-1} z_k^* + z_N^*}, \ \ i=0, \dots, N
\end{equation}
where $z_0^*$, $z_k$ and $z_N$ have to be substituted from Eq.(\ref{zz1})

\begin{figure}[!htb]
\centering
\includegraphics[scale=.75]{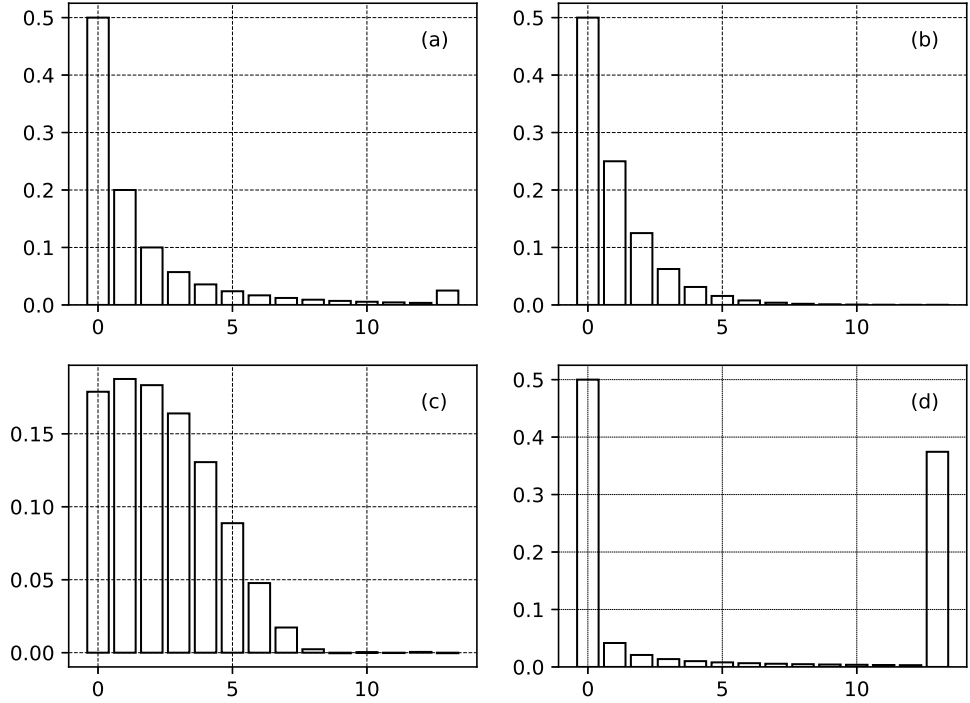}	
\caption{Several distributions connected to the substance $y^*$.
Figure (a): $f_{y,i}=0.002+0.001i$, $\gamma_{y,i}=0.002$. Figure (b):
$f_{y,i}=0.001$, $\gamma_{y,i}=0.001$. Figure (c): $f_{y,i}=0.0015-0.002i$, 
$\gamma_{y,i}=0.0001(i+0.3)$. Figure (d): $f_{y,i}=0.001$, $\gamma_{y,i}=0.001
+ 0.01i$. The profiles of the distributions depend much on the values of parameters.  }	
\end{figure}
Figure 1 shows several profiles of the distribution of the substance $Y$
connected to stationary regime of motion of substances in the channel. Figs.
(a) and (d) demonstrate the effect of accumulation of the substance $Y$
in the last node of the channel. Thus the distribution of the substance can have
long tail for some values of the parameters of the channel but for 
another values of the parameter the long tail can be missing - Fig.(c).
In addition the concentration of the substance along the nodes of the
channel can be different. The substance can be concentrated in the
entry node of the channel - Figs. (a) and (b) or the substance can be
concentrated in the first nodes of the channel - Fig. (c). Finally there
are regimes of flow in which the substance $Y$ is concentrated in the
entry node of the channel and in the last node of the channel - Fig. (d).
	\section{Concluding remarks}
Above we study the motion of reacting substances in a simple
channel of a network (the channel contains only one arm). The study is 
based on a discrete - time model of the channel. The model is
nonlinear one and in the general case of time-dependent coefficients
numerical solution of the model equations is necessary. For the case
of constant coefficients however there are particular cases where
analytical results can be obtained. One such particular case is 
connected to the stationary regime of motion of the substances through the
nodes of the channel. In this case one can obtain analytical
solution of the model equations but this solution is quite complicated.
In order to illustrate the analytical results that are obtained on the
basis of the discussed model we have simplified the situation further:
one of the substances (the substance $Y$) supports the transformation
of part of the substance $X$ to substance of kind $Z$. For this particular case
the mathematical relationships for the distributions of the substances
are simple enough to be written analytically on a relatively small
amount of a paper sheet space. We note that the obtained distributions 
contain as particular cases the distributions for the case of motion of a
single substance in the channel \cite{vr2} and especially the famous 
distributions of Waring, Yule - Simon and Zipf. The study has 
large potential for extending, e.g., two possibilities for this
are: (i) to the case of channel flow in continuous time
\cite{vvz1}, and (ii)  to the case of channel possessing more than 
one arm \cite{vr1}.

\end{document}